\pdfoutput=1
\documentclass[english]{elsarticle}
\usepackage[T1]{fontenc}
\usepackage[latin9]{inputenc}
\usepackage{units}
\usepackage{amsmath}
\usepackage{graphicx}



\usepackage{babel}
\begin{document}

\begin{frontmatter}{}

\title{Improved Numerical Cherenkov Instability Suppression in the Generalized
PSTD PIC Algorithm}

\author{Brendan B. Godfrey}

\address{University of Maryland, College Park, Maryland 20742, USA}

\address{Lawrence Berkeley National Laboratory, Berkeley, California 94720,
USA}

\author{Jean-Luc Vay}

\address{Lawrence Berkeley National Laboratory, Berkeley, California 94720,
USA}
\begin{abstract}
The family of generalized Pseudo-Spectral Time Domain (including the
Pseudo-Spectral Analytical Time Domain) Particle-in-Cell algorithms
offers substantial versatility for simulating particle beams and plasmas,
and well written codes using these algorithms run reasonably fast.
When simulating relativistic beams and streaming plasmas in multiple
dimensions, they are, however, subject to the numerical Cherenkov
instability. Previous studies have shown that instability growth rates
can be reduced substantially by modifying slightly the transverse
fields as seen by the streaming particles . Here, we offer an approach
which completely eliminates the fundamental mode of the numerical
Cherenkov instability while minimizing the transverse field corrections.
The procedure, numerically computed residual growth rates (from weaker,
higher order instability aliases), and comparisons with WARP simulations
are presented. In some instances, there are no numerical instabilities
whatsoever, at least in the linear regime.\end{abstract}

\end{frontmatter}{}

\section{Introduction}

The Pseudo-Spectral Time Domain (PSTD) Particle-in-Cell (PIC) algorithm
advances Fourier-transformed electromagnetic fields in time according
to the difference equations \citep{Liu1997},
\begin{equation}
\mathbf{E}^{n+1}=\mathbf{E}^{n}-i\mathbf{k}\times\mathbf{B}^{n+\nicefrac{1}{2}}\triangle t-\mathbf{J}^{n+\nicefrac{1}{2}}\triangle t,\label{eq:Eleapfrog}
\end{equation}
\begin{equation}
\mathbf{B}^{n+\nicefrac{3}{2}}=\mathbf{B}^{n+\nicefrac{1}{2}}+i\mathbf{k}\times\mathbf{E}^{n+1}\triangle t.\label{eq:Bleapfrog}
\end{equation}
Simulation particles, of course, require fields and produce currents
in real, not Fourier, space. Consequently, Fourier transforms of both
fields and currents are performed at each time step. The fields and
currents in real space typically are located at the nodes of a regular
multidimensional mesh, with magnetic fields $\mathbf{B}$ and currents
$\mathbf{J}$ offset a half time step from electric fields $\mathbf{E}$. 

The PSTD algorithm has been generalized by the authors in two ways
\citep{Vay2014a,Godfrey2014AAC}. First, the components of $\mathbf{k}$
in Eqs. (\ref{eq:Eleapfrog}) and (\ref{eq:Bleapfrog}) are replaced
by the Fourier transforms of various order finite difference approximations
to spatial derivatives on a grid \citep{Lele1992}, which has the
effect of narrowing the effective width of the particle interpolation
stencils. For instance, $k_{i}$ might be replaced by 
\[
\frac{\sin\left(k_{i}\triangle x_{i}/2\right)}{\triangle x_{i}/2},
\]
yielding a pseudo-spectral realization in k-space of the standard
Finite-Difference Time Domain (FDTD) algorithm. (In this context the
components of $\mathbf{k}$ itself can be viewed as an infinite order
approximation.) Second, the field solver is modified to increase the
relatively small PSTD Courant time step limit by a factor of \emph{N},
an integer. This modification is equivalent mathematically to sub-cycling
the field solver \emph{N} times but is faster computationally. Note
that the generalized PSTD algorithm reduces to Haber's Pseudo-Spectral
Analytical Time-Domain (PSATD) algorithm \citep{HaberICNSP73} in
the limit of infinite \emph{N}. Both the details of and the motivation
for the generalized PSTD algorithm, dubbed the Pseudo-Spectral Arbitrary
Order Time Domain (PSAOTD) algorithm, are described thoroughly in
\citep{Vay2014a}. 

The numerical Cherenkov instability \citep{godfrey1974numerical,godfrey1975canonical}
is observed commonly in PIC simulations of relativistic particle beams
and streaming plasmas; \emph{e.g.}, \citep{VayJCP2011,Spitkovsky:ICNSP2011,Xu2013},
where it can be fast growing and strongly disruptive. At a minimum
the instability increases beam emittance \citep{Cormier-Michel2008}.
It arises from nonphysical coupling between the usual electromagnetic
modes, possible distorted by numerical effects, and spurious beam
modes \citep{godfrey2013esirkepov,Godfrey2013PSATD}. The dispersion
relation describing the numerical Cherenkov instability has the general
form,
\begin{multline}
C_{0}+\omega_{p}^{2}\sum_{m_{z}}C_{1}\csc\left[\left(\omega-k_{z}^{\prime}v\right)\frac{\Delta t}{2}\right]+\omega_{p}^{2}\sum_{m_{z}}\left(C_{2x}+\gamma^{-2}C_{2z}\right)\csc^{2}\left[\left(\omega-k_{z}^{\prime}v\right)\frac{\Delta t}{2}\right]\\
+\gamma^{-2}\omega_{p}^{4}\left(\sum_{m_{z}}C_{3z}\csc^{2}\left[\left(\omega-k_{z}^{\prime}v\right)\frac{\Delta t}{2}\right]\right)\left(\sum_{m_{z}}C_{3x}\csc\left[\left(\omega-k_{z}^{\prime}v\right)\frac{\Delta t}{2}\right]\right)=0,\label{eq:drformfull}
\end{multline}
where the $C_{i}$ are complicated functions of the instability frequency
$\omega$, wave numbers $\mathbf{k}$, time step $\Delta t$, cell
size, and alias index $m_{z}$. The particular form of the $C_{i}$
is determined by the details of the numerical algorithm, given in
\citep{Vay2014a} for PSAOTD. The relativistic beam itself is characterized
by its normalized energy $\gamma$, axial velocity $v=\left(1-\gamma^{-2}\right)^{\nicefrac{1}{2}}$,
and normalized relativistic plasma frequency $\omega_{p}$. Of particular
importance are the resonances at $\omega=k_{z}^{\prime}v$, with alias
wave numbers $k_{z}^{\prime}=k_{z}+m_{z}\,2\pi/\triangle z$. (The
beam propagates along the \emph{z}-axis.) Each alias can trigger an
instability, with the most rapid growth typically occurring for $m_{z}=0,-1$.
High order interpolation, say cubic, significantly reduces growth
rates associated with higher order aliases. Although peak growth rates
typically occur at large wave numbers, nontrivial growth rates can
occur at quite small wave numbers, especially for $m_{z}=0$ \citep{godfrey2013esirkepov,Godfrey2013PSATD}.
The nonresonant $m_{z}=0$ instability has two branches \citep{Godfrey2013PSATD,Yu2015},
the primary branch associated with the $C_{2x}$ term of the dispersion
relation, and the secondary branch with the $C_{1}$ term. The former
occurs over a wide range of wave numbers, while the latter occurs
at $k_{z}\approx\nicefrac{\pi}{2\,\triangle z}$ and small $k_{x}$,
if at all. See Fig. \ref{fig:m0branches}. Unless otherwise noted,
parameters for these and other figures are $\omega_{p}=1$, $\gamma=130$,
and $\Delta x=\Delta z=0.3868$. The k-space grid is 65x65, corresponding
to a simulation grid of 128x128.

Until recently, the numerical Cherenkov instability was ameliorated
by a combination of digital filtering, numerical damping, higher order
interpolation (typically cubic), and astute parameter choices \citep{VayJCP2011,Yu2014}.
Papers within the past few years have offered alternative options,
often entailing minor corrections to the transverse electric and magnetic
fields as seen by the particles \citep{Vay2014a,Godfrey2014AAC,Godfrey2013PSATD,Godfrey2014PSATD-TPS}.
Specifically, transverse electric and magnetic fields as they are
interpolated to the particles are multiplied by $\mathbf{k}$-dependent
correction factors $\Psi_{E}$ and $\Psi_{B}$ that vary as $1+\mathcal{O}\left(k^{2}\right)$
for small $\mathbf{k}$. The distinction between such correction factors
and the usual digital filters is made clear in \citep{Godfrey2013PSATD,GodfreyJCP2014}.
The particular analytical correction factors presented in these references
are quite effective at ameliorating the first branch of the $m_{z}=0$
instability but ineffective at ameliorating the second, for reasons
presented later in this article.

As an alternative, this paper numerically computes correction factors
$\Psi_{E}$ and $\Psi_{B}$ just sufficient to completely eliminate
both branches of the $m_{z}=0$ numerical Cherenkov instability. The
process is straightforward in principle. The $m_{z}=0$ dispersion
function is akin to a fifth order polynomial with real coefficients.
If it crosses the \emph{$\omega$}-axis in five places (i.e., five
real roots), all roots are stable. However, if parameters are varied
such that the curve crosses the axis in only three places or one,
then the dispersion function has two or four complex roots, respectively,
the remaining roots being real. The transition occurs where the curve
becomes tangential to the \emph{$\omega$}-axis, \emph{i.e.}, where
both the dispersion function and its derivative with respect to $\omega$
vanish. This determines $\Psi_{E}$ and $\Psi_{B}$. In fact, it can
be shown that the $m_{z}=0$ portion of Eq. (\ref{eq:drformfull})
can be represented as a fifth order polynomial, although this is not
required for the argument of this paragraph to be true.

Fig. \ref{fig:DisFun}, computed for $\nicefrac{v\,\triangle t}{\triangle z}=0.9$
and $k_{z}=k_{x}=5.20$, is illustrative. The left plot displays the
dispersion function, Eq. (\ref{eq:drformfull}) with $m_{z}=0$ only,
for the complete frequency range, $-\nicefrac{\pi}{\triangle t}<\omega<\nicefrac{\pi}{\triangle t}$.
Shown are the results for the uncorrected case, $\Psi_{E}=\Psi_{B}=1$
(labeled Base), for the $C_{2x}$ correction factors described in
\citep{Vay2014a} (labeled $C_{2x}$), and for the numerically determined
optimal correction factors (labeled Opt). (The $C_{2x}$ correction
factors are chosen so that the term $C_{2x}$ vanishes at $\omega=k_{z}v$).
The three curves are essentially indistinguishable on this scale,
with one crossing at the far left and two or four crossings at the
right. The right plot displays a blow-up of the critical frequency
range. Each curve has one crossing in this range, and a second crossing
is off-scale to the right. However, the Opt curve is seen also to
be tangent to the \emph{$\omega$}-axis at a point. Hence, it is stable,
and the other two cases are not. The corresponding $m_{z}=0$ instability
growth rates are 0.0749, 0.0014, and 0 for Base, $C_{2x}$ , and Opt.

The next section provides sample numerical correction factors and
the procedure used to obtain them. The third section then provides
corresponding numerical growth rates from Eq. (\ref{eq:drformfull}),
along with corroborating instability growth rates from WARP two-dimensional
simulations \citep{Warp}. A short concluding section completes the
paper.

The analytical and numerical linear growth rate analyses were performed
using \emph{Mathematica} \citep{Mathematica10}.

\section{Numerical Procedure for Obtaining $\boldsymbol{\mathbf{\Psi}}_{\mathbf{E}}$
and $\boldsymbol{\Psi}_{\mathbf{B}}$ }

The factors $\Psi_{E}$ and $\Psi_{B}$ appear linearly in the $m_{z}=0$
dispersion function, obtained from Eq. \ref{eq:drformfull} by omitting
all but the $m_{z}=0$ terms in the summations, and in its derivative
with respect to $\omega$. As a consequence, the dispersion function
and its derivative together can be inverted readily to obtain $\Psi_{E}$
and $\Psi_{B}$ as functions of the tangent point, $\omega$, although
the actual expressions are quite complicated. 

\begin{figure}
\begin{centering}
\includegraphics[scale=0.45]{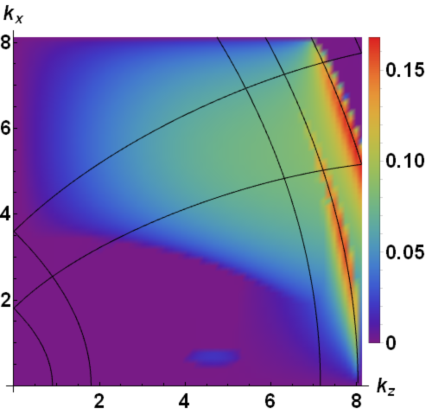}
\par\end{centering}

\protect\caption{\label{fig:m0branches}Growth rates and resonance curves from PSAOTD
dispersion relation for $m_{z}=\left[-2,\,+2\right]$, $v\triangle t/\triangle z=0.9$,
and $N=4$. The secondary $m_{z}=0$ branch, although slower growing
and confined to a small region in \emph{k}-space, is inconveniently
located at small $k_{x}$. }

\end{figure}

\begin{figure}
\begin{centering}
\includegraphics[scale=0.38]{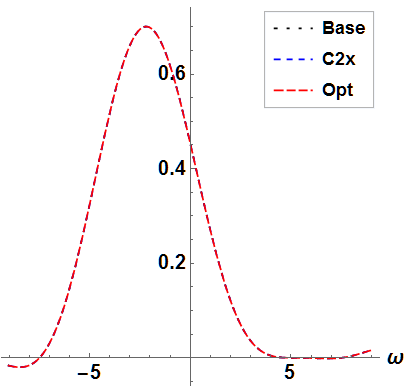}\includegraphics[scale=0.37]{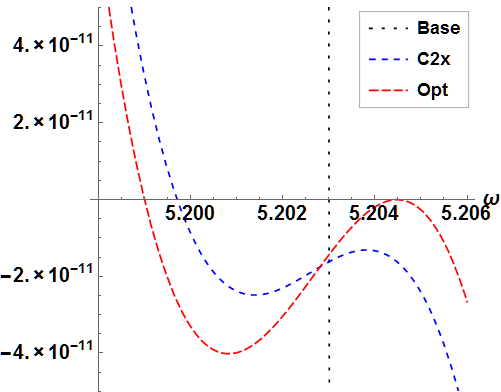}
\par\end{centering}

\protect\caption{\label{fig:DisFun}Dispersion function for Base, $C_{2x}$, and Opt
correction factors. Left, entire range of frequencies (curves indistinguishable
at this scale). Right, narrow range of frequencies near second crossing
point. Parameters are as in Fig. \ref{fig:m0branches}, with $k_{z}=k_{x}=5.20$.}

\end{figure}

\begin{figure}

\begin{centering}
\includegraphics[scale=0.31]{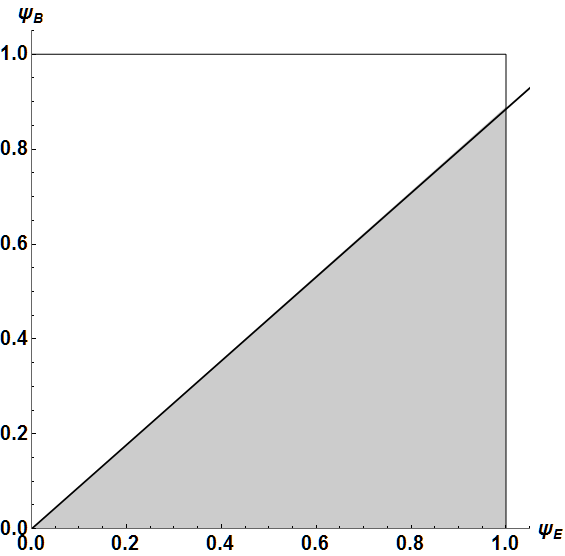}\includegraphics[scale=0.31]{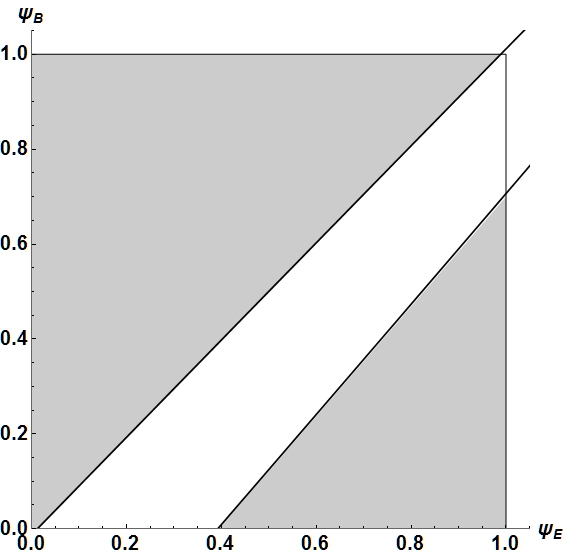}\protect\caption{\label{fig:ParamPlot}$\Psi_{B}-\Psi_{E}$ parametric plots for the
parameters of Fig. \ref{fig:m0branches}; left, $k_{z}=5.20,\: k_{x}=5.20$,
and right, $k_{z}=4.696,\: k_{x}=0.635$. Shaded regions are stable.}

\par\end{centering}

\end{figure}

Fig. \ref{fig:ParamPlot} (left) is a typical parametric plot of $\Psi_{B}$
vs. $\Psi_{E}$ as $\omega$ is varied. Its parameters are identical
to those of Fig. \ref{fig:DisFun}. The region below the diagonal
curve is stable, above it unstable. (Despite appearances, it bends
slightly and does not pass through the origin.) Although any choice
of $\Psi_{B}$ and $\Psi_{E}$ along or below this curve stabilizes
the $m_{z}=0$ mode, a pair close to (1,1) is to be preferred for
minimizing numerical impact on the physics to be simulated. Further
constraining the choice by $\Psi_{B}\leq1$, $\Psi_{E}\leq1$ suggests
the following procedure:
\begin{enumerate}
\item If numerical instability growth is zero at (1,1), set $\Psi_{B}=1$,
$\Psi_{E}=1$.
\item Otherwise, search the lines $\Psi_{B}=1$ and $\Psi_{E}=1$ for the
point nearest (1,1) at which the dispersion function and its derivative
simultaneously vanish. Usually, that point will be on the $\Psi_{E}=1$
line, as illustrated in the left plot of Fig. \ref{fig:ParamPlot}.
\item Otherwise, search for the point nearest (1,1) at which the dispersion
function and its derivative simultaneously vanish in the interior
of the $0<\Psi_{B}<1\;\land\;0<\Psi_{E}<1$ region. It is fortunate
that this situation is rare, because such points are expensive to
find computationally and easy to miss.
\item Otherwise, set $\Psi_{B}=\Psi_{E}=0$. The numerical instability cannot
be eliminated, short of excising that portion of $\mathbf{k}$-space.
\end{enumerate}
For completeness, Fig. \ref{fig:ParamPlot} (right) provides the parametric
plot of $\Psi_{B}$ vs. $\Psi_{E}$ for $k_{z}=4.697,\: k_{x}=0.635$,
which lies at the center of the second branch of the $m_{z}=0$ numerical
Cherenkov instability in Fig. \ref{fig:m0branches}. Stable regions
lie above the upper diagonal curve and below the lower diagonal curve.
Only a very small reduction from unity in $\Psi_{E}$ is sufficient
to eliminate the numerical instability.

\begin{figure}
\begin{centering}
\includegraphics[scale=0.41]{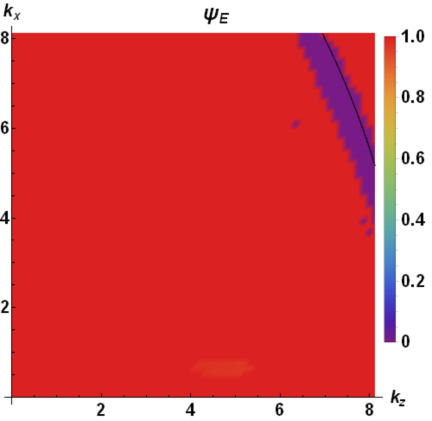}\includegraphics[scale=0.41]{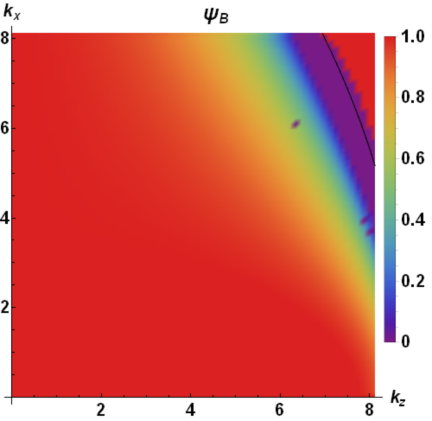}
\par\end{centering}

\protect\caption{\label{fig:factors}$\Psi_{E}$ (left) and $\Psi_{B}$ (right) correction
terms for the parameters of Fig. \ref{fig:m0branches}. Superimposed
at the far upper right of each plot is the $m_{z}=0$ resonance curve.}

\end{figure}

Plots of $\Psi_{E}$ and $\Psi_{B}$ for the parameters of Fig. \ref{fig:DisFun}
are shown in Fig. \ref{fig:factors} with the $m_{z}=0$ resonance
curve superimposed. $\Psi_{E}$ is equal to unity everywhere but (1)
in the small region, barely visible at the bottom center of the plot,
where the second branch of the $m_{z}=0$ numerical Cherenkov instability
appears in Fig. \ref{fig:m0branches}, (2) near the resonance curve,
where $\Psi_{E}=0$, and (3) at three isolated points where $\Psi_{E}=0$
also. $\Psi_{E}\simeq0.97$ for the second branch of the $m_{z}=0$
numerical instability. In contrast, $\Psi_{B}<1$ everywhere except
those $\mathbf{k}$ values where the $m_{z}=0$ growth rate is zero
anyway. Note that $1-\Psi_{B}\simeq0$ over a wide range of small
$\mathbf{k}$.

The three isolated $\Psi_{E}=\Psi_{B}=0$ points appearing in both
plots are associated with a weak instability predicted by Birdsall
and Langdon in Problem 5-9a of their well known text \citep{BirdsallLangdon}.
Because this instability has a very narrow bandwidth in k-space, it
appears only where the instability band happens to overlap nodes on
the k-space numerical grid. Analysis of this Courant-like instability
for the generalized PSTD algorithm is provided in \citep{Vay2014a},
and analysis plus growth rates for the PSATD algorithm in \citep{Godfrey2014PSATD-TPS}.
This instability appears to be unimportant in practice except, perhaps,
at low $\gamma$.

\begin{figure}

\begin{centering}
\includegraphics[scale=0.41]{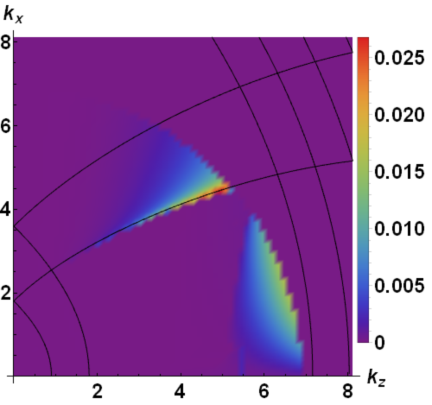}\includegraphics[scale=0.41]{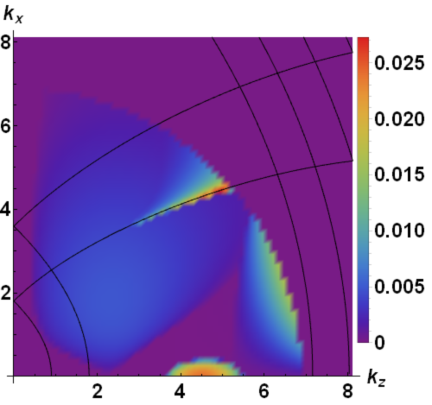}\protect\caption{\label{fig:ebcor4}Growth rates for Fig. (\ref{fig:m0branches}) parameters,
optimal (left) or $C_{2x}$ (right) correction factors, and $\alpha=0.85$.}

\par\end{centering}

\end{figure}

Applying the correction factors of Fig. \ref{fig:factors} for the
parameters used to obtain Fig. \ref{fig:m0branches} completely eliminates
$m_{z}=0$ numerical instability growth but has minimal effect on
higher order aliases except for $\mathbf{k}$-space regions where
$\Psi_{E}$ or $\Psi_{B}$ are much less than unity. Hence, we choose
to truncate fields in $\mathbf{k}$-space according to 
\begin{equation}
k>\alpha\min\left[\frac{\pi}{\triangle z},\frac{\pi}{v\,\triangle t}\right]\label{eq:clip}
\end{equation}
in order to minimized peak growth rates of the first branch of the
$m_{z}=-1$ instability. Fig. \ref{fig:ebcor4} (left) displays the
resulting growth rates for $\alpha=0.85$. Truncating $\mathbf{k}$-space
according to Eq. (\ref{eq:clip}) reduces the peak growth rate from
0.168 to 0.071, and the Fig. \ref{fig:factors} correction factors
further reduce it to 0.026. Most of the residual growth is associated
with the second $m_{z}=-1$ branch. The second branch of the $m_{z}=0$
numerical Cherenkov instability is absent, as expected. In contrast,
Eq. (\ref{eq:clip}) with $C_{2x}$ correction factors reduces peak
growth comparably, but with nontrivial growth rates spread over a
larger range in $\mathbf{k}$-space; see Fig. \ref{fig:ebcor4} (right).
In particular, it causes the second branch of the $m_{z}=0$ numerical
instability to shift toward $k_{x}=0$ but does nothing to eliminate
it. This is because the topology of the $\Psi_{B}$ vs. $\Psi_{E}$
parametric plot changes rapidly for very small $k_{x}$ near this
second instability branch. As a consequence, slightly reducing $\Psi_{E}$
there actually can create an instability where none otherwise exists.

\section{Numerical Instability Growth Rates}

\begin{figure}
\begin{centering}
\includegraphics[scale=0.48]{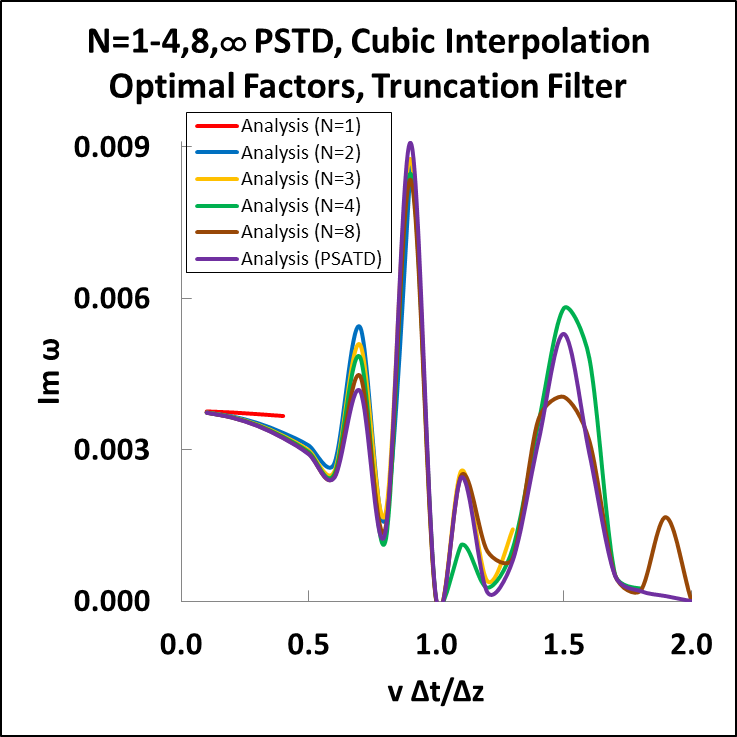}\includegraphics[scale=0.48]{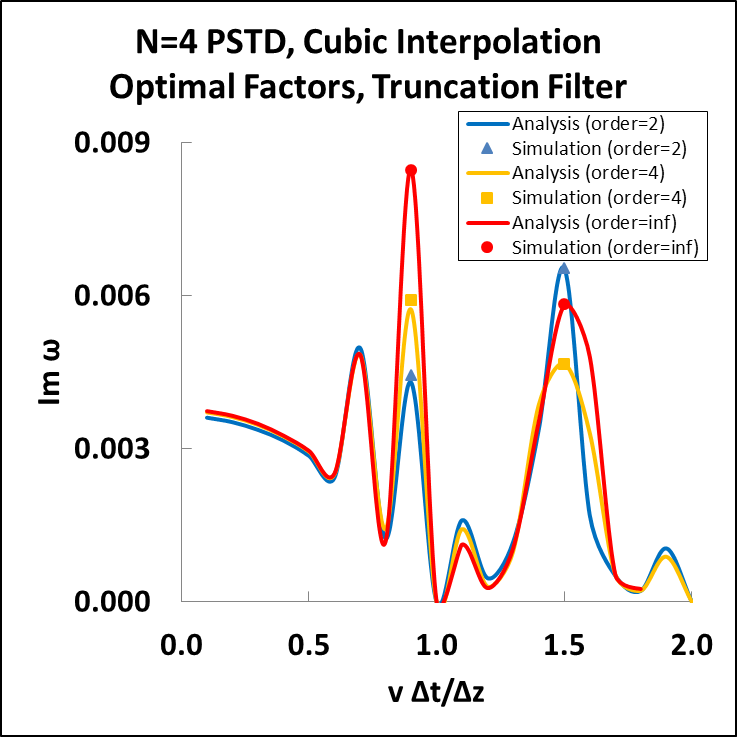}\protect\caption{\label{fig:peakgrowth}Peak numerical instability growth rates \emph{vs}.
$\nicefrac{v\,\triangle t}{\triangle z}$ with for various \emph{N}
(left) and for various order finite difference approximations to $\mathbf{k}$
(right); $\alpha=0.6$.}

\par\end{centering}

\end{figure}

Although $\mathbf{k}$-space truncation with $\alpha=0.85$ yields
reasonably effective instability suppression, $\alpha=0.60$ is much
better. Additionally, choosing $\alpha=0.60$ better accommodates
comparisons with earlier papers \citep{Godfrey2014PSATD-TPS,Godfrey2014AAC,Vay2014a}.
Fig. \ref{fig:peakgrowth} (left) depicts peak growth rates for $N=1,\,2,\,3,\,4,\,8,\,\infty$,
the last being PSATD. Peak growth rates are relatively insensitive
to \emph{N}. As always, the rapid variation of peak growth with $\nicefrac{v\,\triangle t}{\triangle z}$
reflects the rapidly changing locations of higher order resonances
on the $\mathbf{k}$-space grid. Note that the largest growth rates
are associated with $\nicefrac{v\,\triangle t}{\triangle z}$=0.9
, used as the sample case in the preceding Section. No growth whatsoever
is predicted for $\nicefrac{v\,\triangle t}{\triangle z}=1.0$, because
all aliases coincide there. Peak growth rates for $N=4$ and various
order finite difference approximations to $\mathbf{k}$ are shown
in Fig. \ref{fig:peakgrowth} (right). Results are similar to those
in the left plot. Superimposed on the growth rate curves are results
from several WARP simulations \citep{Warp}. Agreement is very good.
In addition, a simulation was run to $t=10000$ for $\nicefrac{v\,\triangle t}{\triangle z}=1.0$
with no sign of instability growth. However, at still larger values
of time, instability growth at a rate of about 0.003 developed at
wave numbers consistent with the second branch of the $m_{z}=0$ numerical
Cherenkov instability. Perhaps, a slight change in the beam distribution
function over time moved this instability above threshold. If so,
reducing $\Psi_{E}$ very slightly there might prevent this very late
time growth.

\begin{figure}
\begin{centering}
\includegraphics[scale=0.48]{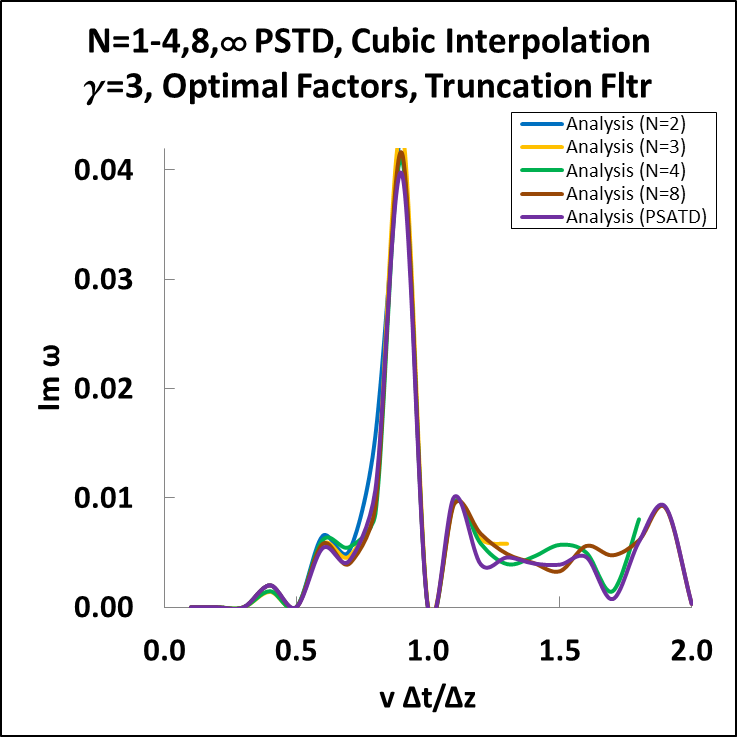}\protect\caption{\label{fig:Peak-numerical-instability-3}Peak numerical instability
growth rates \emph{vs}. $\nicefrac{v\,\triangle t}{\triangle z}$
with $\gamma=3$ and various \emph{N}; $\alpha=0.8$.}

\par\end{centering}

\end{figure}

The procedure introduced in this paper remains robust at low $\gamma$.
Fig. \ref{fig:Peak-numerical-instability-3} presents peak growth
rates for $\gamma=3$, $\alpha=0.8$, $N=2,\,3,\,4,\,8,\,\infty$,
and infinite order. Results are comparable to the best obtained earlier;
see Fig. 7 of \citep{Godfrey2014PSATD-TPS}. Both here and in \citep{Godfrey2014PSATD-TPS},
$E_{z}$ is offset by one-half cell to suppress the usual quasi-electrostatic
numerical instability that occurs at low $\gamma$ \citep{Birdsall1980,lewis1972variational,Langdon1973energy}.
The peak in Fig. \ref{fig:Peak-numerical-instability-3} at $\nicefrac{v\,\triangle t}{\triangle z}=0.9$
is larger than that in Fig. \ref{fig:peakgrowth} only because $\alpha$
is larger.

\section{Conclusion}

New field correction factors have been derived that completely eliminate
the $m_{z}=0$ numerical Cherenkov instability for the generalized
PSTD algorithm, including the PSATD algorithm. When combined with
a sharp cutoff digital filter at large $\mathbf{k}$, these correction
factors reduce peak growth rates to less (often much less) than 0.01
of the beam's relativistic plasma frequency. These coefficients are
optimal in the sense that they differ from unity only slightly over
a broad portion of $\mathbf{k}$-space while eliminating both $m_{z}=0$
numerical instability branches. A disadvantage from the implementation
perspective is that the coefficients must be computed numerically
for each set of simulation parameters. Software for doing so is available
in \emph{Mathematica} CDF format \citep{WolframCDF} at http://hifweb.lbl.gov/public/BLAST/Godfrey/.

\section*{Acknowledgment}

We thank Irving Haber for suggesting this collaboration and for helpful
recommendations. We also are indebted to David Grote for assistance
in using the code WARP. This work was supported in part by the Director,
Office of Science, Office of High Energy Physics, U.S. Dept. of Energy
under Contract No. DE-AC02-05CH11231 and the US-DOE SciDAC ComPASS
collaboration, and used resources of the National Energy Research
Scientific Computing Center.

This document was prepared as an account of work sponsored in part
by the United States Government. While this document is believed to
contain correct information, neither the United States Government
nor any agency thereof, nor The Regents of the University of California,
nor any of their employees, nor the authors makes any warranty, express
or implied, or assumes any legal responsibility for the accuracy,
completeness, or usefulness of any information, apparatus, product,
or process disclosed, or represents that its use would not infringe
privately owned rights. Reference herein to any specific commercial
product, process, or service by its trade name, trademark, manufacturer,
or otherwise, does not necessarily constitute or imply its endorsement,
recommendation, or favoring by the United States Government or any
agency thereof, or The Regents of the University of California. The
views and opinions of authors expressed herein do not necessarily
state or reflect those of the United States Government or any agency
thereof or The Regents of the University of California.

\section*{References}

\bibliographystyle{elsarticle-num}
\bibliography{C:/Users/Brendan/Documents/LyX/Godfrey,C:/Users/Brendan/Documents/LyX/Biblio_JCP_Godfrey}

\clearpage{}
\clearpage{}
\clearpage{}
\clearpage{}
\clearpage{}
\clearpage{}
\clearpage{}

\end{document}